\documentclass[aps, pra, 10pt, twocolumn, superscriptaddress, floatfix]{revtex4-2}
\usepackage[english]{babel}

\usepackage{graphicx}
\usepackage{amsmath}
\usepackage{amssymb}
\usepackage{physics}
\usepackage{nicefrac}
\usepackage{bbold}
\usepackage{braket}
\usepackage{ragged2e}
\usepackage{multirow}
\setcitestyle{numbers}
\usepackage{framed}
\usepackage{bm}
\usepackage{mathtools}

\usepackage[dvipsnames]{xcolor}
\usepackage{hyperref}
\hypersetup{
  colorlinks   = true, 
  urlcolor     = MidnightBlue, 
  linkcolor    = MidnightBlue, 
  citecolor   = Maroon 
}

\renewcommand{\arraystretch}{2.0}

\begin{document}

\title{Complementary 3D color codes for transversal quantum logic}

\author{Friederike Butt}
\affiliation{Institute for Quantum Information, RWTH Aachen University, Aachen, Germany}
\affiliation{Institute for Theoretical Nanoelectronics (PGI-2), Forschungszentrum J\"{u}lich, J\"{u}lich, Germany}

\author{Luis Colmenarez}
\affiliation{Institute for Quantum Information, RWTH Aachen University, Aachen, Germany}
\affiliation{Institute for Theoretical Nanoelectronics (PGI-2), Forschungszentrum J\"{u}lich, J\"{u}lich, Germany}

\author{Erik Weilandt}
\affiliation{Technische Universität München, Munich, Germany}

\author{Tom Peham}
\affiliation{Technische Universität München, Munich, Germany}

\author{Robert Wille}
\affiliation{Technische Universität München, Munich, Germany}

\author{Markus Müller}
\affiliation{Institute for Quantum Information, RWTH Aachen University, Aachen, Germany}
\affiliation{Institute for Theoretical Nanoelectronics (PGI-2), Forschungszentrum J\"{u}lich, J\"{u}lich, Germany}

\begin{abstract}
Transversal logical gates provide a direct route to fault-tolerant quantum computation, but the Eastin--Knill theorem forbids a universal transversal gate set within a single quantum error-correcting code. We propose a \emph{hybrid} architecture based on the tetrahedral three-dimensional color code and its Hadamard-transformed counterpart, which we call the \emph{H-tetrahedral} code. The two encodings support complementary transversal non-Clifford operations. Combined with bitwise Hadamard transformations that switch between the two encodings and a one-way transversal logical CNOT from the tetrahedral code to the H-tetrahedral code, these operations realize an almost-universal transversal logical gate set that enables both the creation of entanglement and logical states with magic. We complete a universal gate set through a pieceably fault-tolerant round-robin construction of a logical controlled-$Z$ gate between two H-tetrahedral codes. This logical entangling gate is interleaved with reduced-overhead Steane-type syndrome extraction using logical two-dimensional color-code auxiliary qubits. Our construction provides a new route toward implementing classically hard-to-simulate quantum algorithms where magic and most entangling operations are transversal while the resource overhead is concentrated in a small number of non-transversal Clifford entangling operations. 

\end{abstract}

\maketitle

\section{Introduction}

Quantum error correction (QEC) is a fundamental ingredient for achieving fault-tolerant (FT) quantum computation \cite{gottesman1997stabilizer,terhal2015quantum}. QEC works by distributing logical information across multiple noisy qubits so that the encoded logical qubits become more resilient to noise, while the redundant degrees of freedom enable the detection and correction of errors. Several recent works have demonstrated the advantages of QEC in suppressing errors on logical qubits \cite{google2025quantum,lacroix2025scaling,caune2024demonstrating,putterman2024hardware,reichardt2024demonstration,bluvstein2024logical,reichardt2024fault,bluvstein2025fault,mathiot2026benchmarking}. In addition, logical gate operations must be implemented in such a way that the encoded information is transformed as intended while preventing errors from spreading uncontrollably during gate execution. 
An implementation that inherently suppress the propagation of errors with nearly no overhead are \emph{transversal logical gates}, which act independently on local disjoint sets of physical qubits. 
Nevertheless, the Eastin--Knill theorem~\cite{eastin2009restrictions} forbids the existence of a transversal \emph{universal} logical gate set on a single QEC code. As a consequence, universal FT quantum computation requires alternative mechanisms for implementing the missing logical operations~\cite{campbell_roads_2017}.

One prominent strategy is the preparation of magic states, which enable the application of non-transversal logical gates through gate teleportation~\cite{gottesman1999demonstrating,chou2018deterministic}. Magic state injection has been demonstrated on various hardware platforms~\cite{yamamoto2026quantum, butt2026demonstration, mayer2024benchmarking, postler2022demonstration, sales2025experimental,
bluvstein2025fault}. 
In this paradigm, the central challenge is the FT preparation of sufficiently high-fidelity magic states. The original proposal for preparing these high-fidelity resource states, magic-state distillation~\cite{bravyi2005universal, sales2025experimental}, combines many noisy non-FT magic states to produce fewer, higher-fidelity ones, albeit at a substantial resource cost. More recently, magic-state cultivation~\cite{gidney2024magic, sahay2025fold} has been proposed and demonstrated~\cite{rosenfeld2025magic} as a potentially less resource-intensive alternative while still enabling high-quality magic-state preparation. Another promising approach is code switching~\cite{bombin2015gauge,anderson2014fault,kubica2015universal, butt2024fault}, where the encoded state is transferred between complementary QEC codes so that the logical gate that is missing for universality can be implemented transversally in the new encoding. Code switching has recently been demonstrated on trapped-ion~\cite{pogorelov2025experimental, daguerre2025experimental} and neutral-atom platforms~\cite{bluvstein2025fault}. 
Nevertheless, realizing universal logical gate sets with lower resource overhead remains one of the central challenges in fault-tolerant quantum computation.

\begin{figure*}[!tb]
	\centering
	\includegraphics[width=\linewidth]{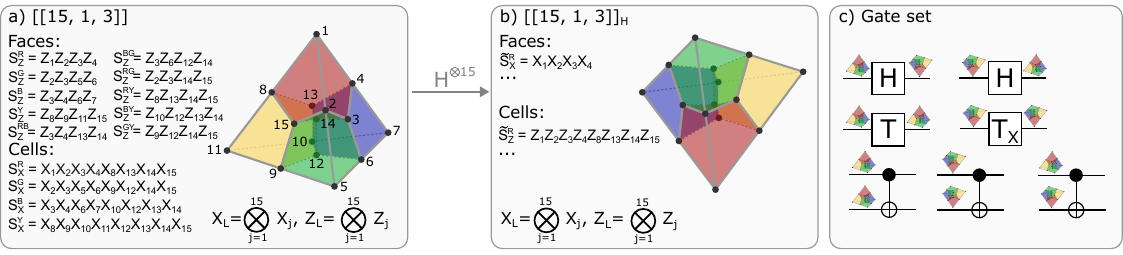}
	\caption{\justifying \textbf{The $[[15, 1, 3]]$ code and its Hadamard-transformed 
    version. }(a) The $X$-stabilizer generators of the $[[15, 1, 3]]$ tetrahedral color code are supported on the red (R), blue (B), green (G) and yellow (Y) cell formed by eight qubits each. Ten independent $Z$-stabilizers are defined on weight-4 faces within the tetrahedron. (b) The support of $X$- and $Z$-stabilizers is interchanged on the Hadamard-transformed tetrahedral code $[[15, 1, 3]]_\mathrm{H}$: $X$-stabilizers are supported by faces and $Z$-stabilizers are supported by cells. (c) The joint gate set of both code variants comprises the logical CNOT between the same code types, the T gate on the $[[15, 1, 3]]$ code and the T$_X$ gate on the $[[15, 1, 3]]_\mathrm{H}$ code. Applying $\Tilde{H} = \bigotimes_{j=1}^{15} H_j$ implements the desired logical operation but changes the encoding. Additionally, a one-way CNOT gate can be implemented transversally if the control qubit is encoded in a $[[15, 1, 3]]$ code and the target qubit in a $[[15, 1, 3]]_\mathrm{H}$ code. }
	\label{fig:tetrahedral_color_code}
\end{figure*}

In this work, we propose a new route toward universal FT quantum computation based on transversal operations across distinct but closely related QEC codes. Specifically, we consider the tetrahedral code~\cite{bombin2015gauge,kubica2015universal} together with its Hadamard-transformed counterpart, which we refer to as the H-tetrahedral code. The latter is obtained by applying a Hadamard gate to every physical qubit of the original tetrahedral code, thereby transforming the stabilizer structure accordingly. 
The tetrahedral and H-tetrahedral codes support transversal $\frac{\pi}{4}$-rotations along the $Z$- and $X$-axis, respectively. 
As a result, each code individually enables logical rotations about one Pauli axis, bringing the combined architecture closer to a universal logical gate set~\cite{kitaev1997quantum,dawson2005solovay}.
Furthermore, the two codes can be entangled through a one-way transversal CNOT gate~\cite{heussen2025efficient, tan2025single}, where the tetrahedral code acts as the control and the H-tetrahedral code as the target. The transversal CNOT in the reversed direction, from the H-tetrahedral code as control to the tetrahedral code as target logical qubit, 
is not available, so the resulting two-logical-qubit gate set is not universal by itself. Nevertheless, this structure still enables the implementation of previously inaccessible non-Clifford and entangling logical operations using transversal gates only. Importantly, these transversal operations exist for arbitrary code distances~\cite{bombin2007topological,bombin2013self}, making the 
available transversal gate set inherently scalable. 
Consequently, such an architecture could potentially simplify the implementation of hard-to-simulate quantum algorithms while relying on transversal logical operations. 
A natural application of this setting is the implementation of the class of instantaneous quantum polynomial-time (IQP) circuits~\cite{shepherd2009temporally}: these circuits are not universal but can be implemented with transversal gates~\cite{hangleiter2025fault, paletta2024robust} and certain instances are believed to be hard to simulate on classical machines~\cite{bremner2017achieving}. 
Transversality of logical operations makes the hybrid architecture particularly attractive for hardware platforms that support highly parallel operations, such as neutral-atom architectures~\cite{bluvstein2024logical,evered2023high}, since many transversal layers can be implemented in parallel.

To complete the universal gate set, we propose implementing the missing CNOT operation through a round-robin construction~\cite{yoder2016universal}, which incorporates intermediate rounds of partial syndrome extraction during the execution of the logical gate. This approach falls under the framework of pieceable fault tolerance~\cite{yoder2016universal, ryan2022implementing}. Moreover, we reduce the overhead associated with syndrome extraction by employing Steane-type extraction with 2D color-code logical ancilla qubits instead of 3D color-code logical ancilla states. 
We explicitly confirm fault tolerance of our constructions for the smallest error-correcting code instances by means of numerical simulations. 

The structure of the paper is as follows. We first review the construction of three-dimensional (3D) tetrahedral color codes and their native logical gates in Sec.~\ref{sec:almost_univ_gate_set}. We then introduce a \emph{hybrid} architecture that combines a tetrahedral and a H-tetrahedral color code and discuss the available transversal gate set. In Sec.~\ref{sec:compelting_univ_gate_set}, we then complete a FT universal gate set by constructing a round-robin entangling gate with compact intermediate stabilizer extraction. We analyze the performance of this gate and conclude in Sec.~\ref{sec:conclusion}.

\section{Transversal gates with complementary encodings}\label{sec:almost_univ_gate_set}

\subsection{Tetrahedral color code}
\begin{figure*}[!tb]
	\centering
	\includegraphics[width=\linewidth]{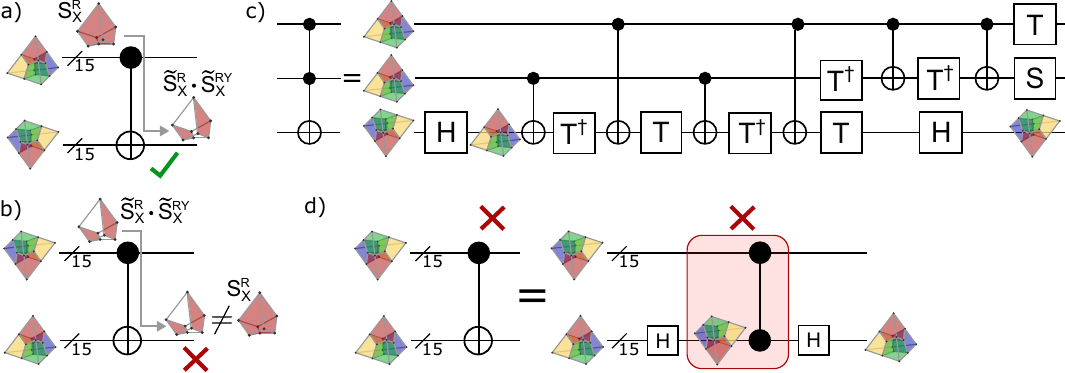}
	\caption{\justifying \textbf{Transversal operations in the hybrid architecture. }(a) If the control qubit is encoded in the $[[15, 1, 3]]$ code and the target in the Hadamard-transformed $[[15, 1, 3]]_\mathrm{H}$ code, cell-like stabilizers propagate to face-like stabilizers. This operation preserves the codespace because both control and target are a $+1$-eigenstate of the $X$-cells. The transversal CNOT gate in the reverse direction (b) does not preserve the codespace, because an $X$-face propagates from control to the target code that is not an eigenstate of the respective face-like stabilizer. (c) The hybrid architecture admits a transversal Toffoli gate if the two control qubits are encoded in $[[15, 1, 3]]$ codes and the target qubit is encoded in the $[[15, 1, 3]]_\mathrm{H}$ code. (d) We decompose the CNOT gate that does not preserve the codespace into a C$Z$ gate and Hadamard operations. The forbidden transversal logical entangling gate therefore corresponds to a $\Tilde{\mathrm{C}Z}$ gate between two $[[15, 1, 3]]_\mathrm{H}$ codes. }
	\label{fig:stabilizer_Prop_one_way_CNOT}
\end{figure*}

The tetrahedral color code~\cite{bombin2007topological, kubica2015unfolding, butt2026decoding} is defined on a four-valent, four-colorable lattice in which each vertex is connected to four neighboring vertices. Qubits are associated with the vertices, while subsets of qubits define three-dimensional regions called cells and two-dimensional planar regions called faces.
Four colors, red (R), green (G), blue (B) and yellow (Y) are assigned to cells within the 3D structure in such a way that two cells of one color do not share any vertex in the lattice. 
Additionally, each face is labeled by the combined colors of the two cells it separates. For instance, the interface between a red cell and a green cell is labeled red--green. Typically, each cell supports an $X$-type stabilizer, while each face supports a Z-type stabilizer.
In the following, we exemplarily consider the smallest error-correcting distance $d=3$ code instance: the $[[15,1,3]]$ code encodes $k=1$ logical qubit in $n= 15$ physical qubits and any single error can be corrected. 
As summarized in Fig.~\ref{fig:tetrahedral_color_code}, the $X$-type stabilizer generators are defined on the cells, such that
\begin{equation}
\begin{aligned}
& S_X^\mathrm{R} = &X_1 X_2 X_3 X_4 X_8 X_{13} X_{14} X_{15}, \\
& S_X^\mathrm{G} = &X_2 X_3 X_5 X_6 X_9 X_{12} X_{14} X_{15}, \\
&S_X^\mathrm{B} =  & X_3 X_4 X_6 X_7 X_{10} X_{12} X_{13} X_{14}, \\
& S_X^\mathrm{Y} =  &  X_8 X_9 X_{10} X_{11} X_{12} X_{13} X_{14} X_{15},
\end{aligned}
\end{equation}
where $X_i$ and $Z_i$ denote the Pauli operators on the $i$th physical qubit. 
The faces of the tetrahedron support the stabilizers
\begin{equation}
\begin{aligned}
S_Z^\mathrm{R}   &= Z_1 Z_2 Z_3 Z_4,      &\quad S_Z^\mathrm{BG} &= Z_3 Z_6 Z_{12} Z_{14}, \\
S_Z^\mathrm{G}   &= Z_2 Z_3 Z_5 Z_6,      &\quad S_Z^\mathrm{RG} &= Z_2 Z_3 Z_{14} Z_{15}, \\
S_Z^\mathrm{B}   &= Z_3 Z_4 Z_6 Z_7,      &\quad S_Z^\mathrm{RY} &= Z_8 Z_{13} Z_{14} Z_{15}, \\
S_Z^\mathrm{Y}   &= Z_8 Z_9 Z_{11} Z_{15},&\quad S_Z^\mathrm{BY} &= Z_{10} Z_{12} Z_{13} Z_{14}, \\
S_Z^\mathrm{RB}&= Z_3 Z_4 Z_{13} Z_{14},&\quad S_Z^\mathrm{GY} &= Z_9 Z_{12} Z_{14} Z_{15}.
\end{aligned}
\end{equation}
The outer faces on the boundary of the tetrahedron correspond to stabilizers $S^\mathrm{R}_Z,S^\mathrm{G}_Z,S^\mathrm{B}_Z,S^\mathrm{Y}_Z$. The interfaces between cells of two colors support the remaining $Z$-type stabilizers labeled by the two colors of the adjacent cells. For example, $S^\mathrm{BG}_Z$ corresponds to the face between the blue and green cells. While only a single $Z$-error is correctable in the $[[15, 1, 3]]$ code, up to three $X$-errors can be corrected. In the following, we distinguish between the overall distance $d$ and the effective distance with respect to a specific Pauli-error type, namely $d = d_Z = 3$ and $d_X = 7$. In the following, we denote logical gate operations $O$ that preserve the codespace by $\overline{O}$. 
The logical Pauli operators $\overline{X}$ and $\overline{Z}$ of the tetrahedral color code can be implemented by applying the respective Pauli operation to all 15 physical qubits. 
The $[[15, 1, 3]]$ code supports a transversal $\overline{T} = e^{-i \pi \overline{Z}/8}$ gate that can be implemented by applying~\cite{kubica2015universal, bombin2007topological}
\begin{equation}\label{eq:t_gate_def}
\overline{T} = \bigotimes_{j=1}^{15} T^{\dag}_j. 
\end{equation}
One can verify that Eq.~\eqref{eq:t_gate_def} implements a valid logical $\overline{\mathrm{T}}$ gate by checking that the codespace is preserved and that the logical generators are transformed accordingly as $\overline{\mathrm{T}} \overline{X} \overline{\mathrm{T}^{\dagger}} = (\overline{X} + \overline{Y})/\sqrt{2}$ and $\overline{\mathrm{T}} \overline{Z} \overline{\mathrm{T}^{\dagger}} = \overline{Z}$.

\begin{figure*}[!tb]
	\centering
	\includegraphics[width=1\linewidth]{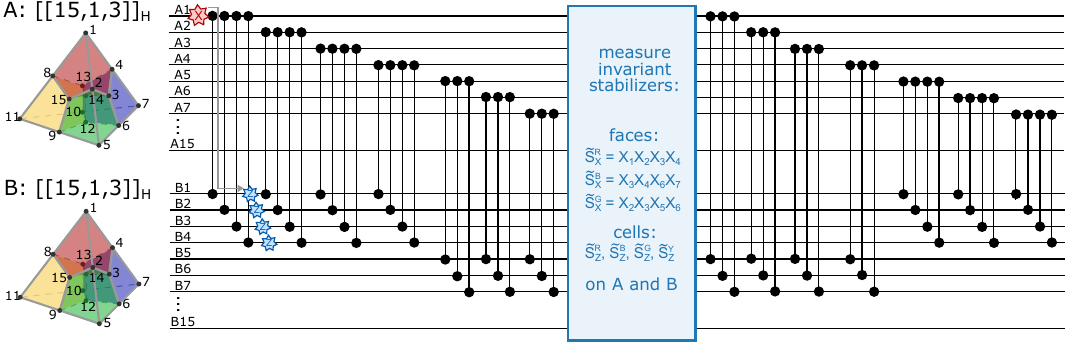}
	\caption{\justifying \textbf{Round-robin C$Z$ gate on the rotated tetrahedral $[[15, 1, 3]]_\mathrm{H}$ code}. The round-robin gate applies seven layers of permuted transversal C$Z$ gates on qubits 1--7 of code blocks A and B. 
    Since physical C$Z$ gates commute, we can rearrange the gate ordering, as illustrated in App.~Fig.~\ref{fig:CZ_rearrangement}. In doing so, the protocol becomes symmetric between the two registers, which allows us to use the same lookup tables for both logical code blocks. 
    The subset of \emph{invariant} stabilizers remain unchanged throughout the whole protocol. This subset of stabilizers is extracted in the middle of the circuit (blue box). Specifically, we measure the red, blue and green $X$-faces and three out of four $Z$-cells of both encoded blocks by means of Steane-type stabilizer extraction, and the yellow cell stabilizer $\Tilde{S}_Z^\mathrm{Y}$ of both registers using a flag-based circuit. A single $X$-error (red) may propagate to four $Z$-errors (blue) on the complementary register. A weight-four $Z$-error would not be correctable on either code block. However, with the information about the invariant stabilizers in combination with a full round of stabilizer extraction after the full round-robin gate, these propagated high-weight errors remain correctable. }
	\label{fig:round_robin_gate_circuit}
\end{figure*}

The logical $\overline{S}$ gate can be implemented by applying $\overline{\mathrm{T}}$ twice and the transversal logical $\overline{\mathrm{CNOT}}$ is available as in any CSS code~\cite{gottesman1997stabilizer}. 
Consequently, also the controlled-S gate and the three-qubit CC$Z$ gate are transversal on the tetrahedral color code. The missing gate for universality is the logical Hadamard gate $\overline{H}$. The bitwise application of $\Tilde{H} = \bigotimes_{j=1}^{15} H_j$ transforms the logical operators correctly as $\Tilde{H} \overline{X} \Tilde{H} = \overline{Z}$ and $\Tilde{H} \overline{Z} \Tilde{H} = \overline{X}$, but it does not preserve the codespace, because $X$- and $Z$-stabilizers have different support. In the following, $\Tilde{O}$ denotes a transversal gate operation $O$ that does not implement a codespace-preserving logical operation. 

\subsection{H-tetrahedral color code}

In this section we define the 
Hadamard-transformed tetrahedral code, dubbed H-tetrahedral code $[[15,1,3]]_{\mathrm{H}}$. 
We obtain this code by taking the initial tetrahedral $[[15, 1, 3]]$ code and applying $\Tilde{H} = \bigotimes_{j=1}^{15} H_j$, such that $X\rightarrow Z$ and  $Z\rightarrow X$ on each stabilizer and logical operator, as illustrated in Fig.~\ref{fig:tetrahedral_color_code}(b). 
The H-tetrahedral code has a transversal logical gate set about an axis, which is rotated with respect to the tetrahedral code, such that the rotation $\overline{T}_X = \Tilde{H} \overline{T} \Tilde{H} = e^{- i \pi \overline{X}/8}$ is now transversal.  
As a consequence, the transversal logical Clifford operation $\overline{\sqrt{X}} = \overline{T}^2_X$ is available.
Interestingly both the $[[15, 1, 3]]$ and the $[[15, 1, 3]]_\mathrm{H}$ code are related by a bitwise Hadamard transformation, enabling arbitrary single-qubit logical rotations up to this Hadamard transformation. 
In the next section we discuss how this can be exploited by designing an architecture with both code variants. 

\begin{figure*}[!tb]
	\centering
	\includegraphics[width=1\linewidth]{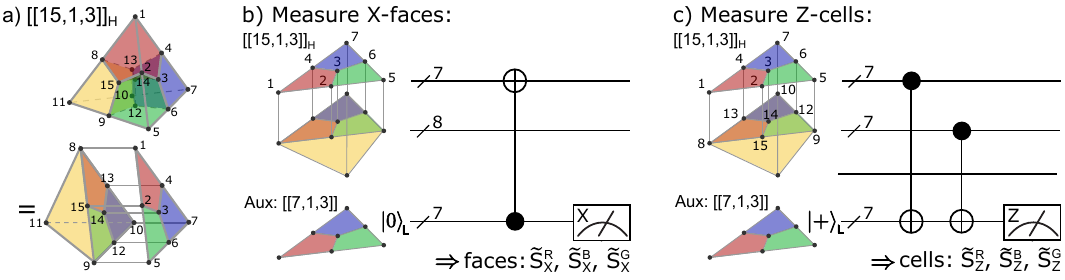}
	\caption{\justifying \textbf{Fault-tolerant stabilizer extraction on tetrahedral 3D color codes. }(a) A 3D tetrahedral color code can be viewed as layers of 2D color codes, similarly to stacked color codes~\cite{jochym2016stacked}. (b) The red, green and blue $X$-stabilizers, $S_X^\mathrm{R}, S_X^\mathrm{B}, S_X^\mathrm{G}$ can be measured fault-tolerantly using a variant of Steane's method for FT syndrome extraction~\cite{steane1997active}. We fault-tolerantly prepare a logical auxiliary qubit in $|\overline{0}\rangle$ of a 2D color code by verifying the prepared logical state after running a unitary encoding~\cite{goto2016minimizing}, as shown in App.~Fig.~\ref{fig:encoding_steane_code}. In a second step, we apply a transversal CNOT gate between the top layer of the 3D code instance (qubit 1--7) and the logical auxiliary qubit. Measuring the auxiliary register in the $X$-basis reveals the $X$-syndrome. (c) Similarly, we measure three cell-like stabilizers using a logical auxiliary qubit encoded in a 2D color code. We apply two layers of transversal CNOT gates between two layers of the stacked 3D code and the auxiliary 2D code instance. The fourth stabilizer supported by the yellow cell can either be measured with the same construction but considering different pairs of layers within the tetrahedron, or using a standard flag-based circuit for measuring a $Z$-stabilizer~\cite{chao2018fault}. }
	\label{fig:stabilizer_extraction}
\end{figure*}

\subsection{Hybrid architecture and one-way CNOT}

We now consider two logical qubits, one encoded in the original tetrahedral code and the other in the H-tetrahedral code. 
On the original tetrahedral $[[15, 1, 3]]$ code, we can implement a transversal T gate, on the H-tetrahedral $[[15, 1, 3]]_\mathrm{H}$ code, we can realize a transversal T$_X$ gate, as summarized in Fig.~\ref{fig:tetrahedral_color_code}(c). The transversal operation $\Tilde{H}$ implements the logical Hadamard gate and simultaneously changes the encoding from \hbox{$[[15, 1, 3]] \longrightarrow [[15, 1, 3]]_\mathrm{H}$} and vice versa. 

A \emph{one-way} transversal CNOT gate can be implemented by applying a bitwise CNOT operation, with the tetrahedral code serving as the control and the H-tetrahedral code as the target.
In this direction, only cell-like stabilizers propagate to face-like stabilizers. A cell corresponds to a pair of faces, such that the codespace is preserved under this operation, as illustrated in Fig.~\ref{fig:stabilizer_Prop_one_way_CNOT}(a). 
The transversal one-way CNOT gate also realizes the desired logical operation because $\overline{X}_\mathrm{control}$ and $\overline{Z}_\mathrm{target}$ transform accordingly to $\overline{X}_\mathrm{control} \overline{X}_\mathrm{target}$ and $\overline{Z}_\mathrm{control}\overline{Z}_\mathrm{target}$, respectively. Unfortunately the same bitwise CNOT operation with inverted target and control does not preserve the codespace, which limits the transversal CNOT realization to only one direction. Hence this architecture supports transversal CNOTs between the same kind of logical qubits, $X$- and $Z$-rotations on either kind of qubit and \emph{some} transversal entangling operation between the two kind of logical qubits. Therefore the transversal logical gate set is \emph{almost} universal but still able to generate entanglement, due to available CNOT operations, and magic~\cite{campbell2017roads, howard2017application}, due to the different non-Clifford gates on each kind of logical qubit.

The transversal logical gate set of the hybrid architecture is not universal, in agreement with the Eastin-Knill theorem~\cite{eastin2009restrictions}. A universal gate set requires the missing entangling CNOT operation to realize arbitrary operations on any of the two logical qubits. 
In our construction, arbitrary logical single-qubit operations are generally not available in either of the two encoding, as each code only supports transversal rotations about one axis. 
For example, magic state injection on the H-tetrahedral code using a magic state prepared on the tetrahedral code requires the missing CNOT operation. 
Therefore, arbitrary single logical qubit operations on the H-tetrahedral code are not enabled by the transversal operations in the hybrid architecture. 

The proposed near-universal hybrid architecture is scalable to higher-distance 3D tetrahedral color codes, as these codes preserve the same stabilizer and logical-operator group structures across all code distances. 
Although the transversal gate set is not universal, it supports \emph{some} entangling and non-Clifford operations, 
enabling the implementation of transversal sub-blocks with few physical operations within a larger algorithm. 
For example, the Toffoli gate is transversal for a specific configuration of encodings, as illustrated in Fig.~\ref{fig:stabilizer_Prop_one_way_CNOT}(c).

\section{Completing a universal gate set}\label{sec:compelting_univ_gate_set}

\subsection{Pieceable fault-tolerant $\overline{\mathrm{C}Z}$ gate}
In order to complete a universal gate set, we resort to the concept of pieceable fault tolerance~\cite{yoder2016universal}. The key idea is that even if an entire logical circuit is not FT, parts of the circuit may be. Additional syndrome extractions are inserted in between FT sub-circuits to maintain fault tolerance for the entire protocol. 
Pieceable FT gates have been shown to complete universal logical gate sets in different QEC codes~\cite{yoder2017universal, lin2020concatenated, takagi2017error}. 
In our hybrid architecture, one direction of a logical entangling gate is missing for achieving universality, where the control qubit is encoded in a $[[15, 1, 3]]_\mathrm{H}$ and the target in a $[[15, 1, 3]]$ code. 
We decompose the missing CNOT gate into a C$Z$ gate with adjacent Hadamard operations on the target qubit as shown in Fig.~\ref{fig:stabilizer_Prop_one_way_CNOT}(d), such that the missing entangling operation corresponds to a C$Z$ gate between two $[[15, 1, 3]]_\mathrm{H}$ codes. In this section, we construct a FT implementation of this C$Z$ gate based on pieceable FT round-robin gates~\cite{yoder2016universal, chao2018fault}. 

Formally, the round-robin construction of an $h$-qubit gate $U$ disjoint sets of qubits $\Lambda_1,\ldots \Lambda_h$ is given by~\cite{yoder2016universal}
\begin{align}
    \prod_{j_1 \in \Lambda_1} \ldots \prod_{j_h \in \Lambda_h} U(j_1, \ldots j_h). 
\end{align}
In our case, $h=2$ for the two-qubit gate $U = \mathrm{C}Z$ and $\Lambda_1 = 1, 2, 3, 4, 5, 6, 7$, $\Lambda_2 = 1', 2', 3', 4', 5', 6', 7'$. 
Specifically, the round-robin $\overline{\mathrm{C}Z}$ gate between two $[[15, 1, 3]]_\mathrm{H}$ codes applies seven layers of transversal C$Z$ gates between two code blocks, as shown in App.~Fig.~\ref{fig:CZ_rearrangement}. Each layer consists of seven physical C$Z$ gates that are applied between the physical qubits that support one representation of $\overline{Z}$, for example the right red--blue--green side of the tetrahedron containing qubits 1--7. Between each layer, the target qubits are permuted: in the first layer, we apply C$_{A1}Z_{B1}\,$C$_{A2}Z_{B2}$...C$_{A7}Z_{B7}$. In the second layer, we continue by applying C$_{A1}Z_{B2}\,$C$_{A2}Z_{B3}$...C$_{A7}Z_{B1}$, and so on. In the last layer, we finally apply C$_{A1}Z_{B7}\,$C$_{A2}Z_{B1}$...C$_{A7}Z_{B6}$. We rearrange the order of physical C$Z$ gates such that the gate implementation is symmetric between the two registers, as illustrated in Fig.~\ref{fig:round_robin_gate_circuit}. 

By construction, the stabilizers are preserved for the full protocol because all stabilizers overlap at an even number of sites with the support of the round-robin gate,~i.e., with qubits 1--7. However, during the protocol, the stabilizer group changes, and only a subset of stabilizers, called \emph{invariant} stabilizers, remains unchanged throughout the whole protocol. 
All $Z$-stabilizers, $\Tilde{S}_Z^\mathrm{R}, \Tilde{S}_Z^\mathrm{B}, \Tilde{S}_Z^\mathrm{G}, \Tilde{S}_Z^\mathrm{Y}$, of both code registers remain unchanged because they commute with all C$Z$ gates. Additionally, three $X$-type stabilizers on each encoded block, $\Tilde{S}_X^\mathrm{R}, \Tilde{S}_X^\mathrm{B}, \Tilde{S}_X^\mathrm{G}$, are preserved. Their support fully overlaps with the physical qubits participating in the round-robin gate and, therefore, in total overlap at an even number of sites with the support of the logical gate. 
The naive round-robin construction is not FT since a single $X$-error on one qubit in one register, as illustrated in Fig.~\ref{fig:round_robin_gate_circuit},  propagates to a logical error on the complementary register. We insert one reduced round of stabilizer extraction in the middle of the circuit, as shown in Fig.~\ref{fig:round_robin_gate_circuit}. Here, we only need to measure the invariant stabilizers, which are the red, blue and green $X$-faces of both registers, $\Tilde{S}_X^\mathrm{R}, \Tilde{S}_X^\mathrm{B}, \Tilde{S}_X^\mathrm{G}$, and all four cells, $\Tilde{S}_Z^\mathrm{R}, \Tilde{S}_Z^\mathrm{B}, \Tilde{S}_Z^\mathrm{G}, \Tilde{S}_Z^\mathrm{Y}$, of both code blocks. After executing the full circuit, we run one full round of stabilizer extraction and, together with the information about the intermediate invariant stabilizers, one can correct any single fault that may have propagated through the circuit.

\subsection{Reduced-overhead stabilizer extraction with 2D logical ancillas}

The stabilizers of the 3D color code must be measured fault-tolerantly to preserve fault tolerance in our logical entangling gate. To this end, we use a modified version of Steane-type syndrome extraction~\cite{steane1997active}. Steane's method for FT stabilizer measurements consists of three steps. First, a logical auxiliary qubit encoded in the same code as the data register is prepared fault-tolerantly. Second, a transversal CNOT gate is applied to transfer error information from the data register to the auxiliary qubit register. Finally, the auxiliary logical qubit is measured destructively in an appropriate basis to extract the syndrome.
Steane's method has been used to implement repeated rounds of QEC using various codes on different platforms~\cite{reichardt2024fault, bluvstein2025fault, bluvstein2024logical, huang2024comparing, postler2022demonstration}. 

Instead of encoding the logical auxiliary qubit in another 3D color code, we prepare the logical auxiliary qubit in a 2D color code. 
We identify each layer of the 3D tetrahedron with a 2D code instance, as illustrated in Fig.~\ref{fig:stabilizer_extraction}(a). 
To measure three $X$-stabilizers supported on faces within the tetrahedron of the distance-3 3D color code, we first fault-tolerantly prepare an auxiliary register in $|\overline{0}\rangle$ of a $[[7, 1, 3]]$ code as defined in App.~\ref{app:steane_code_definition}. Here, we first prepare the logical state by means of a unitary encoding circuit, followed by a measurement of a suitable logical operator that heralds the presence of errors that otherwise lead to a logical failure~\cite{goto2016minimizing}. The circuit for FT state preparation on the $[[7, 1, 3]]$ code is shown in App.~Fig.~\ref{fig:encoding_steane_code}.
After initializing a logical auxiliary state, we apply a transversal CNOT gate between the seven data qubits that support the stabilizers to be measured and the logical auxiliary qubit, as shown in Fig.~\ref{fig:stabilizer_extraction}(b). Similarly to the one-way CNOT gate discussed above, this transversal CNOT preserves the codespace~\cite{heussen2025efficient, tan2025single}. It also copies $Z$-errors from data qubits to the auxiliary register, where $Z$- errors can be detected through an auxiliary qubit readout in the $X$-basis. As expected from Steane-type syndrome extraction, single $X$-errors on the auxiliary logical qubit that may backpropagate to the data-qubit register remain correctable. 
During the intermediate round of invariant $X$-stabilizer measurements within our round-robin gate, we only need to measure the red, blue and green $X$-faces $\Tilde{S}_X^{R}, \Tilde{S}_X^{B}, \Tilde{S}_X^{G}$. Hence, a single instance of Steane-type stabilizer extraction is sufficient and only a single logical auxiliary qubit is needed. In contrast, when measuring all ten independent $X$-stabilizers, one requires at least four instances because only three $X$-stabilizers can be measured at once. 

\begin{figure}[!tb]
	\centering
	\includegraphics[width=1\linewidth]{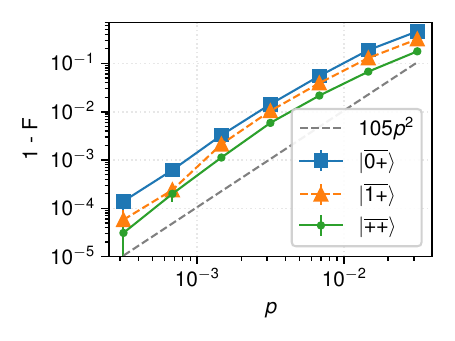}
	\caption{\justifying \textbf{Logical state fidelities for the fault-tolerant round-robin $\overline{\mathrm{C}Z}$ gate. } The $\overline{\mathrm{C}Z}$ gate is run on two $[[15, 1, 3]]_\mathrm{H}$ codes which do not admit a transversal implementation of this operation. The gate construction is fully symmetric between the two logical qubits. The logical state fidelities are determined as described in App.~\ref{app:logical_fidelity_calculation} for two logical qubit input states $|\overline{0+}\rangle, |\overline{1+}\rangle$, and $|\overline{++}\rangle$. The gray dashed line corresponds to the approximated logical infidelity of a transversal $\overline{\mathrm{CNOT}}$ gate, assuming that any two faults contribute to the logical infidelity. }
	\label{fig:results_RR_CZ_gate}
\end{figure}

We can use a similar approach for measuring $Z$-stabilizers supported on weight-8 cells, as shown in Fig.~\ref{fig:stabilizer_extraction}(c). As a first step, a single logical auxiliary qubit encoded in a $[[7, 1, 3]]$ code is prepared fault-tolerantly in $|+\rangle_\mathrm{L}$ as described in App.~\ref{app:steane_code_definition}. We then apply two layers of transversal CNOT gates. The first transversal CNOT gate acts on pairs of qubits on one layer of the 3D code (qubits 1--7) and the seven qubits of the logical auxiliary qubit. This operation alone does not preserve the codespace, because, for example, a face-like $Z$-stabilizer of the auxiliary $[[7, 1, 3]]$ is mapped to the tetrahedral $[[15, 1, 3]]$ code which is not a $+1$-eigenstate of this $Z$-stabilizer. 
To restore the codespace, we apply the second transversal CNOT gate between the next layer of the 3D code (qubits 8, 9, 10, 12, 13, 14, 15) and the logical auxiliary code. The face-like $Z$-stabilizers of the auxiliary $[[7, 1, 3]]$ are now mapped twice to the tetrahedral $[[15, 1, 3]]$ code onto two opposing faces. Two opposing faces make a cell, for example the red face supported by qubits 1--4 and the red--yellow face supported by qubits 8, 13, 14, 15 together form the red cell. In total, the stabilizers of the 3D color code are therefore again preserved. A final measurement of the auxiliary register in the $Z$-basis reveals three of the cell-like $Z$-stabilizers. 
A single $Z$-error on the auxiliary qubit can backpropagate to two data qubits; however, the resulting error remains correctable by the $[[15,1,3]]_\mathrm{H}$ code as we have an effective distance $d_Z = 7$ for $Z$-errors.
To measure all four cells of the $[[15, 1, 3]]_\mathrm{H}$ code, we have to repeat this full block at least twice as only three $Z$-stabilizers can be measured at once. As a more resource-efficient alternative, we measure the red, blue and green cell using the modified Steane-type stabilizer extraction as illustrated in Fig.~\ref{fig:stabilizer_extraction}(c) once, and then extract the information about the remaining yellow cell using a standard flag-based circuit with two physical auxiliary qubits~\cite{chao2018fault}. 
Since tetrahedral and H-tetrahedral have the same stabilizer structure, the Steane-type syndrome extraction naturally applies to both codes.

\subsection{Numerical simulation}
To verify fault-tolerance of our construction, we deterministically place all single fault configurations within our circuits and confirm that no single fault leads to a logical failure on any of the two logical qubits. 
In addition, we perform Monte Carlo simulations considering single-parameter circuit-level noise, as described in detail in App.~\ref{app:numerical_methods}. Every component in the circuit introduces an error with the same probability $p$. The noisy $\overline{\mathrm{C}Z}$ is run on error-free logical input states but includes the noisy extraction of the invariant stabilizers. Figure~\ref{fig:results_RR_CZ_gate} shows the numerically simulated logical state infidelities of the logical two-qubit state after the application of the round-robin $\overline{\mathrm{C}Z}$ gate between two $[[15, 1, 3]]_\mathrm{H}$ codes on input states $|\overline{0+}\rangle, |\overline{1+}\rangle$, and $|\overline{++}\rangle$. The logical state fidelities are determined as described in App.~\ref{app:logical_fidelity_calculation}. 
The logical error rate scaling $p_\mathrm{L}\sim p^2$ for all three input states signals the successful correction of all single-fault events. For comparison, we estimate the logical infidelity of a transversal $\overline{\mathrm{CNOT}}$ between two $[[15, 1, 3]]_\mathrm{H}$ codes. Under the pessimistic assumption that any two faults on any of the 15 physical CNOT gates leads to a failure, we approximate the logical infidelity of the transversal $\overline{\mathrm{CNOT}}$ to be $1 - F_\mathrm{CNOT} \approx \binom{15}{2} p^2 = 105 p^2$, which is shown in Fig.~\ref{fig:results_RR_CZ_gate}. The approximated infidelity of the transversal $\overline{\mathrm{CNOT}}$ is almost one order of magnitude lower than for the round-robin $\overline{\mathrm{C}Z}$ gate since there are fewer fault locations due to a smaller circuit depth. 

\subsection{Scaling up}
The proposed construction of the logical $\overline{\mathrm{C}Z}$ gate extends naturally to higher-distance codes. 
A H-tetrahedral code with parameters $[[(d^3 + d)/2, 1, d]]$~\cite{hangleiter2025fault, butt2026decoding} has a distance 
\begin{align}
  d_Z = \frac{3d^2 + 1}{4} \propto d^2 > d
\end{align}
for $Z$-type errors, as derived in App.~\ref{app:distance_relation}. 
This distance $d_Z$ determines the support of the logical Pauli operator $\overline{Z}$ and, therefore, the size of the round-robin gate construction. Specifically, $d_Z$ layers of transversal $\mathrm{C}Z$ gates have to be applied that each contain $d_Z$ physical C$Z$ gates. Within the full logical operation, we need to perform $(d -1)/2$ rounds of invariant-stabilizer extraction to maintain fault tolerance for the full protocol.
These FT stabilizer measurements utilize distance-$d$ 2D color code instances encoded in $d_Z$ physical qubits. 
Each round of invariant-stabilizer extraction consists of three layers of transversal C$Z$ gates containing $d_Z$ physical C$Z$ gates. In total, the full logical operation thus requires $\mathcal{O}(d\cdot d_Z^2) = \mathcal{O}(d^5) $ physical C$Z$ gates provided that logical resource states for the stabilizer measurements are available.

\section{Discussion and Outlook}\label{sec:conclusion}

In this work, we have introduced a hybrid architecture based on the tetrahedral color code and its Hadamard-transformed counterpart. By combining these two encodings, we obtain a set of transversal operations that can realize non-Clifford single-qubit rotations and several types of entangling operations. The remaining missing entangling operation can be implemented via a round-robin construction with a compact intermediate stabilizer extraction based on Steane-type stabilizer extraction~\cite{steane1997active}, thereby completing a universal fault-tolerant logical gate set. 

A practical implementation of a larger algorithm in this architecture requires a QEC cycle between logical operations~\cite{aliferis2005quantum, cross2007comparative, gutierrez2015comparison}. 
State-of-the-art methods for constructing fault-tolerant algorithms insert such QEC cycles after every non-transversal gadget and any non-Clifford gate~\cite{zhou2025low, serra2026decoding}. 
The hybrid architecture offers the possibility of generating both entanglement and magic using transversal logical gates only. As a result, QEC does not necessarily need to be performed after every non-Clifford operation, but may instead be inserted only after an optimized number of transversal layers. The optimal frequency of these QEC cycles within a larger algorithm determines the overall resource overhead. Finding the optimal QEC schedule for the presented hybrid color-code architecture for a given method of stabilizer extraction is therefore an important direction for future work.

To fully realize the potential of our hybrid architecture, several important challenges remain to be addressed. In particular, a detailed resource analysis that accounts for the capabilities and constraints of specific hardware platforms will be essential. For example, highly parallel operations available in neutral-atom architectures~\cite{bluvstein2024logical, evered2023high} could substantially reduce the circuit depth, as each layer of the round-robin gate and stabilizer extraction circuits based on Steane-type QEC may be implemented in a single parallelized step. However, the overhead associated with atom shuttling and qubit rearrangement may offset some of these advantages in practice. Incorporating hardware-specific noise models could further improve performance estimates. For instance, biased noise, which naturally arises in trapped-ion systems~\cite{haffner2008quantum}, may be exploited to enhance the performance of the protocol.
A thorough understanding of the trade-offs between performance and resource overhead will be crucial for comparing our approach with other routes to universal fault-tolerant quantum computation, including magic-state injection~\cite{bravyi2005universal, preskill1998reliable}, magic-state distillation~\cite{bravyi2005universal}, magic-state cultivation~\cite{sahay2025fold, gidney2024magic}, and code switching~\cite{anderson2014fault}. Such comparisons will help clarify the regimes in which the proposed hybrid architecture offers practical advantages.

A relevant figure of merit is not only the cost of implementing a non-transversal logical operation, but also the frequency with which such operations occur in fault-tolerant quantum algorithms. In the proposed architecture, logical magic and many entangling operations are available as native transversal gates, while only one of the six possible orientations (three code combinations and two directions) of entangling operations requires a more involved fault-tolerant protocol. This feature distinguishes our approach from architectures in which every non-Clifford gate necessitates a costly magic-state protocol or a code-switching procedure.
More broadly, the proposed architecture suggests a route to universality in which the overhead is concentrated in a small number of non-transversal entangling operations, while logical magic and many entangling gates remain available transversally. An important direction for future work is to minimize the number of costly C$Z$ gates within a larger algorithm and a detailed analysis of the implementation of QEC cycles and decoding techniques. 
A systematic study of compilers achieving such a reduction and of the performance of QEC cycles will help pave the way for universal fault-tolerant quantum computation with low-overhead by combining complementary transversal gate sets with a small number of pieceably fault-tolerant operations.

\section*{Acknowledgements}

This work was funded by the Deutsche 
Forschungsgemeinschaft (DFG, German Research Foundation) through Grant No. 449905436 and Grant No. 563402549. 
This research is also part of the Munich Quantum Valley (K-5 and K-8), which is supported by the Bavarian state government with funds from the Hightech Agenda Bayern Plus. 
The authors acknowledge funding from the European Research Council (ERC) under the European Union's Horizon 2020 program (Grant Agreement No. 101001318) and from the Horizon Europe program under Grant Agreement Number 101114305 ("MILLENION-SGA1" EU Project)
We also acknowledge support by the German Federal Ministry of Research, Technology and Space (BMFTR) as part of the Research Program Quantum Systems, research project 13N17317 (”SQale”),  
the BMFTR project MUNIQC-ATOMS (Grant No. 13N16070), and the Intelligence Advanced Research Projects Activity (IARPA), under the Entangled Logical Qubits program through Cooperative Agreement Number W911NF-23-2-0216.
We furthermore acknowledge support by the Deutsche Forschungsgemeinschaft (DFG, German Research Foundation) under Germany’s Excellence Strategy ‘Cluster of Excellence Matter and Light for Quantum Computing (ML4Q) EXC 2004/1’ 390534769. The views and conclusions contained in this document are those of the authors and should not be interpreted as representing the official policies, either expressed or implied, of IARPA, the Army Research Office, or the U.S. Government. The U.S. Government is authorized to reproduce and distribute reprints for Government purposes notwithstanding any copyright notation herein. We acknowledge computing time provided at the NHR Center NHR4CES at RWTH Aachen University (Project No. p0020074). This is funded by the Federal Ministry of Education and Research and the state governments participating on the basis of the resolutions of the GWK for national high-performance computing at universities. 

\section*{Author contributions:}
F.B. developed the presented protocols and performed all presented numerical simulations. F.B. and L.C. analyzed results and wrote the manuscript, with contributions from all authors. R.W. and M.M. supervised the project. 

\newpage
\appendix
\section{Lookup table for round-robin C$Z$ gate}

Table~\ref{tab:lut_round_robin_intermediate} summarizes the lookup table for the intermediate round of invariant-stabilizer extraction in the middle of the round-robin $\overline{\mathrm{C}Z}$ gate between two qubits encoded in $[[15, 1, 3]_\mathrm{H}$ codes. The lookup table is symmetric between the two registers, so A/B can correspond to the first/second code block or vice versa. Additionally, we track dangerous errors,~i.e. single $X$-errors propagating to three or four $Z$-errors on the complementary qubit register. A single $X$-error on block A is identified by the $Z$-stabilizers of logical qubit A. If the $X$-syndrome on block B in the intermediate invariant-stabilizer extraction does not match any of those listed in Tab.~\ref{tab:lut_round_robin_intermediate}, we count it as a dangerous error. In combination with one full round of stabilizer extraction after the full protocol, these dangerous errors can finally be corrected.

\section{The $[[7, 1, 3]]$ code}\label{app:steane_code_definition}
\begin{figure}[!b]
	\centering
	\includegraphics[width=90mm]{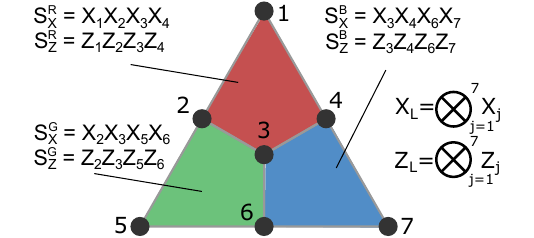}
	\caption{\justifying \textbf{Stabilizer generators and logical operators of the $[[7, 1, 3]]$ code.}  }
	\label{fig:colorcode_standalone}
\end{figure}

\begin{figure*}[!tb]
	\centering
	\includegraphics[width=\linewidth]{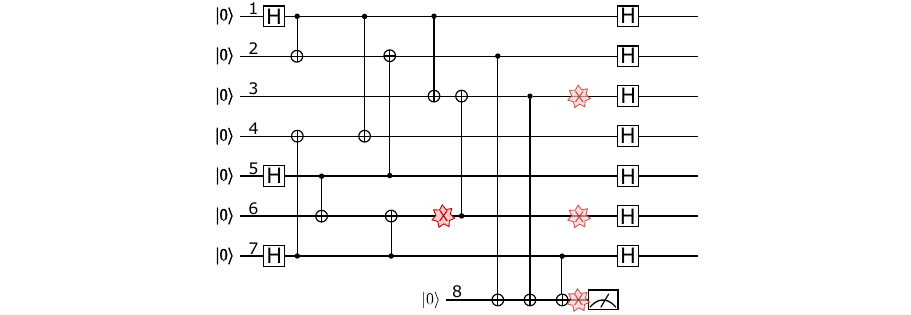}
	\caption{\justifying \textbf{Circuit for the fault-tolerant preparation of a logical state on the $[[7, 1, 3]]$ code~\cite{goto2016minimizing}}. We first initialize $|\overline{0}\rangle$ on the $[[7, 1, 3]]$ non-fault-tolerantly using the first eight CNOT gates. Then, we detect single faults that would otherwise cause a logical failure by measuring the logical operator $Z_{\mathrm{L}} = Z_2 Z_3 Z_7$. Finally, a transversal application of H$_{\mathrm{L}}$ may be used to prepare $|+\rangle_{\mathrm{L}}$. }
	\label{fig:encoding_steane_code}
\end{figure*}

Figure~\ref{fig:colorcode_standalone} summarizes the stabilizers and logical operators of the $[[7, 1, 3]]$ code. This code is the smallest instance of an error-correcting two-dimensional color code~\cite{bombin2006topological} and it has been used in several experimental demonstrations~\cite{goto2016minimizing, daguerre2025experimental, lacroix2025scaling, pogorelov2025experimental, postler2024demonstration, postler2022demonstration}.
Figure~\ref{fig:encoding_steane_code} shows the circuit for FT logical state preparation on the $[[7, 1, 3]]$ code~\cite{goto2016minimizing}. After the unitary encoding, a verification step measures a suitable logical operator that anticommutes with all weight-2 errors that result from a single fault in the unitary circuit. 

\section{C$Z$ gate ordering of round-robin $\overline{\mathrm{C}Z}$ gate on two $[[15, 1, 3]]_\mathrm{H}$ codes}\label{app:CZ_ordering}

\begin{figure*}[!tb]
	\centering
	\includegraphics[width=\linewidth]{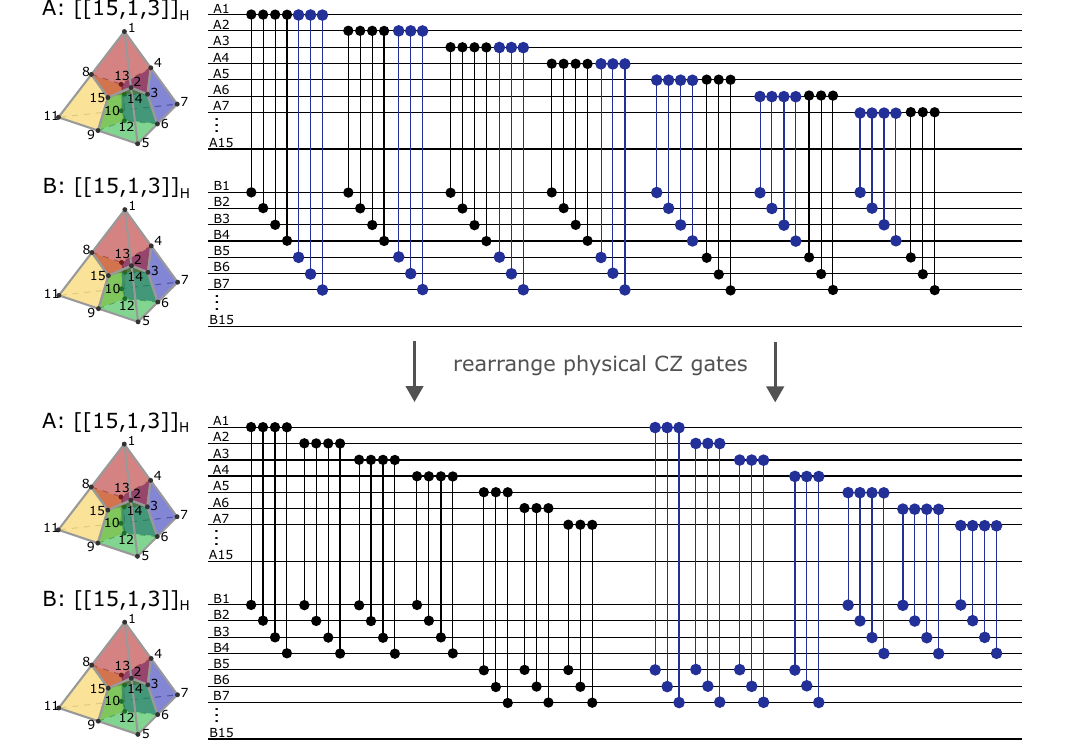}
	\caption{\justifying \textbf{Ordering of physical C$Z$ gates in the round-robin $\overline{\mathrm{C}Z}$ gate.} We change the ordering of physical C$Z$ gates from the naive round-robin implementation (top). Specifically, we move all gates colored in blue to the second half of the protocol (bottom). As a consequence, the lookup tables taking into account the intermediate round of stabilizer extraction in the FT round-robin gate is the same for both logical-qubit registers. }
	\label{fig:CZ_rearrangement}
\end{figure*}

Figure~\ref{fig:CZ_rearrangement} illustrates the ordering of physical C$Z$ gates in the straightforward round-robin gate. We rearrange these gates for convenience of decoding.

\section{Numerical methods}\label{app:numerical_methods}

We simulate noisy quantum circuits by performing Monte Carlo simulations. We use the \textit{Pauli-twirling approximation} (PTA) to model noise on each circuit component as stochastic Pauli channels. 
Each noisy circuit component is modelled by first applying the respective ideal operation, followed by an error $E$ occurring with probability $p$. 
We simulate a depolarizing noise channel after every single- and two-qubit gate
\begin{align}
    \mathcal{E}_1(\rho) &= (1 - p_1)\rho + \frac{p_1}{3} \sum_{j= 1}^3 E^{j}_1 \rho E^{j}_1 \label{eq:depol_single_qubit} \\
    \mathcal{E}_2(\rho) &= (1 - p_2)\rho + \frac{p_2}{15} \sum_{j= 1}^{15}  E_2^{j} \, \rho\, E_2^{j}. \nonumber
\end{align}
Here, an error is applied with probability $p_1$ and $p_2$ from the error sets $E_1$ and $E_2$ defined by
\begin{align}
	E_1 &\in \{ \sigma_k, \forall k \in \{1, 2, 3 \} \} \\
	E_2 &\in \{\sigma_k \otimes \sigma_l, \forall k, l \in \{0, 1, 2, 3 \}  \} \backslash  \{\sigma_0 \otimes \sigma_0 \}, \nonumber
\end{align}
where $\sigma_k$ are the single-qubit Pauli operators \hbox{$\sigma_k = \{I, X, Y, Z \}$ with $k=0, 1, 2, 3$}. 
All qubits are initialized and measured in the $Z$-basis. We simulate faults in these operations by applying $X$-errors after initialization and before measurement, each occurring with probabilities $p_{\mathrm{init}}$ and $p_{\mathrm{meas}}$, respectively. In the presented simulations, we consider a uniform noise model and set $p = p_1 = p_2 = p_{\mathrm{init}} = p_{\mathrm{meas}}$. 
We do not include noise on idling qubits, dephasing, leakage, qubit loss or crosstalk. 

\section{Determination of logical state fidelity}\label{app:logical_fidelity_calculation}

The logical state fidelities shown in Fig.~\ref{fig:results_RR_CZ_gate} are determined for logical input states $|\overline{0+}\rangle, |\overline{1+}\rangle$, and $|\overline{++}\rangle$. We numerically initialize the respective error-free logical input state and then apply a noisy $\overline{\mathrm{C}Z}$ gate as described in Sec.~\ref{sec:compelting_univ_gate_set}. After each run, we measure all qubits destructively in a Pauli basis and average over all runs to determine the expectation values. Specifically, the states $\overline{\mathrm{C}Z}|\overline{0+}\rangle = |\overline{0+}\rangle,\, \overline{\mathrm{C}Z}|\overline{1+}\rangle = |\overline{1-}\rangle$ and $\overline{\mathrm{C}Z}|\overline{++}\rangle$ are stabilized by \{$\overline{Z}_1,\,\overline{X}_2\},\, \{-\overline{Z}_1,\,-\overline{X}_2\}$ and \{$\overline{X_1Z_2}, \overline{Z_1X_2}$\}, respectively. 
The projectors onto the logical two-qubit output states are given
\begin{align}
    P_{\mathrm{C}Z|0+\rangle} &= \frac{1}{2}\left( 1 +  Z_1 \right)\frac{1}{2}\left( 1 +  X_2\right) \\
    P_{\mathrm{C}Z|1+\rangle} &= \frac{1}{2}\left( 1 - Z_1 \right)\frac{1}{2}\left( 1 -X_2 \right) \nonumber\\
    P_{\mathrm{C}Z|++\rangle} &= \frac{1}{2}\left( 1 +  Z_1X_2 \right) \frac{1}{2}\left( 1 + X_1Z_2  \right)\nonumber. 
\end{align}
The logical state fidelities are then determined as
\begin{align}
    F_{\mathrm{C}Z|0+\rangle} &= \frac{1}{4}\left( 1 + \langle Z_1 \rangle + \langle X_2 \rangle + \langle Z_1 X_2 \rangle \right) \\
    F_{\mathrm{C}Z|1+\rangle} &= \frac{1}{4}\left( 1 - \langle Z_1 \rangle - \langle X_2 \rangle + \langle Z_1 X_2 \rangle \right) \nonumber\\
    F_{\mathrm{C}Z|++\rangle} &= \frac{1}{4}\left( 1 + \langle Z_1X_2 \rangle + \langle X_1Z_2 \rangle + \langle Y_1Y_2 \rangle\right)\nonumber. 
\end{align}

\section{Relation between distances $d_X$ and $d_Z$ in the tetrahedral color code}\label{app:distance_relation}

On the H-tetrahedral color code, the effective distance for $X$-errors $d_X = d$ is given by the length of a border which is a string along one edge of the tetrahedral structure~\cite{bombin2007topological}. The effective distance for $Z$-errors is determined by the area of a boundary which is a sheet on one side of the tetrahedron~\cite{bombin2007topological}. We identify a boundary with a 2D color code with a 6.6.6 tiling~\cite{landahl2011fault}. The number of physical qubit of the 2D color code $n_{\mathrm{2D}}$ therefore coincides with the distance $d_Z$ of the H-tetrahedral 3D color code, and the distance of the 2D color code coincides with the distance $d$ of the H-tetrahedral color code. The function $d_Z(d)$ is therefore given by the relation between the number of qubits in a 2D color code and its distance
\begin{align}
    d_Z = n_{\mathrm{2D}} = \frac{3d^2 + 1}{4}.
\end{align}

\begin{table*}[!tb]
\centering
\renewcommand{\arraystretch}{1.5}
\begin{tabular}{c|c|c|c}
$S_{Z,A}^{\mathrm{Y}}$ 
& $(S_{Z,A}^{\mathrm{R}}, S_{Z,A}^{\mathrm{B}}, S_{Z,A}^{\mathrm{G}})$ 
& $(S_{X,B}^{\mathrm{R}}, S_{X,B}^{\mathrm{B}}, S_{X,B}^{\mathrm{G}})$ 
& Correction \\
\hline
\hline
$0$ & $(100)$ & $(000)$ & $X_{A_1}$ \\
$0$ & $(110)$ & $(000)$ & $X_{A_2}$ \\
$0$ & $(111)$ & $(000)$ & $X_{A_3}$ \\
$0$ & $(101)$ & $(000)$ & $X_{A_4}$ \\
$0$ & $(010)$ & $(000)$ & $X_{A_5}$ \\
$0$ & $(011)$ & $(000)$ & $X_{A_6}$ \\
$0$ & $(001)$ & $(000)$ & $X_{A_7}$ \\
\hline
$0$ & $(100)$ & $(100)$ & $X_{A_1} Z_{B_2} Z_{B_3} Z_{B_4}$ \\
$0$ & $(110)$ & $(100)$ & $X_{A_2} Z_{B_2} Z_{B_3} Z_{B_4}$ \\
$0$ & $(111)$ & $(100)$ & $X_{A_3} Z_{B_2} Z_{B_3} Z_{B_4}$ \\
$0$ & $(101)$ & $(100)$ & $X_{A_4} Z_{B_2} Z_{B_3} Z_{B_4}$ \\
\hline
$0$ & $(100)$ & $(010)$ & $X_{A_1} Z_{B_3} Z_{B_4}$ \\
$0$ & $(110)$ & $(010)$ & $X_{A_2} Z_{B_3} Z_{B_4}$ \\
$0$ & $(111)$ & $(010)$ & $X_{A_3} Z_{B_3} Z_{B_4}$ \\
$0$ & $(101)$ & $(010)$ & $X_{A_4} Z_{B_3} Z_{B_4}$ \\
\hline
$0$ & $(100)$ & $(101)$ & $X_{A_1} Z_{B_4}$ \\
$0$ & $(110)$ & $(101)$ & $X_{A_2} Z_{B_4}$ \\
$0$ & $(111)$ & $(101)$ & $X_{A_3} Z_{B_4}$ \\
$0$ & $(101)$ & $(101)$ & $X_{A_4} Z_{B_4}$ \\
\hline
$0$ & $(010)$ & $(010)$ & $X_{A_5} Z_{B_6} Z_{B_7}$ \\
$0$ & $(011)$ & $(010)$ & $X_{A_6} Z_{B_6} Z_{B_7}$ \\
$0$ & $(001)$ & $(010)$ & $X_{A_7} Z_{B_6} Z_{B_7}$ \\
\hline
$0$ & $(010)$ & $(001)$ & $X_{A_5} Z_{B_7}$ \\
$0$ & $(011)$ & $(001)$ & $X_{A_6} Z_{B_7}$ \\
$0$ & $(001)$ & $(001)$ & $X_{A_7} Z_{B_7}$ \\
\hline
$1$ & $(100)$ & any & $X_{A_8}$ \\
$1$ & $(110)$ & any & $X_{A_{15}}$ \\
$1$ & $(111)$ & any & $X_{A_{14}}$ \\
$1$ & $(101)$ & any & $X_{A_{13}}$ \\
$1$ & $(010)$ & any & $X_{A_9}$ \\
$1$ & $(011)$ & any & $X_{A_{12}}$ \\
$1$ & $(001)$ & any & $X_{A_{10}}$ \\
\end{tabular}
\caption{Lookup table for the intermediate correction during the round-robin gate. We denote the measured $Z$-cell syndrome on the first block (A) by $S_{Z,A}$ and the measured $X$-face syndrome on the second block (B) by $S_{X,B}$. The flag $S_{Z,A}^{\mathrm{Y}} = 0$ indicates an error on the Steane face of the tetrahedron, whereas $S_{Z,A}^{\mathrm{Y}} = 1$ indicates an error on the yellow cell. Blocks A and B can be used interchangeably, because our gate construction is symmetric.}
\label{tab:lut_round_robin_intermediate}
\end{table*}

\clearpage
\bibliography{references}

@article{terhal2015quantum,
  title={Quantum error correction for quantum memories},
  author={Terhal, B. M.},
  journal={Rev. Mod. Phys.},
  volume={87},
  number={2},
  pages={307--346},
  year={2015},
  doi={10.1103/RevModPhys.87.307},
  url={https://doi.org/10.1103/RevModPhys.87.307}
}

@misc{gottesman1997stabilizer,
  title={Stabilizer Codes and Quantum Error Correction},
  author={Gottesman, Daniel},
  year={1997},
  howpublished={PhD thesis, California Institute of Technology},
  note={\url{https://thesis.library.caltech.edu/2900/2/THESIS.pdf}}
}

@article{google2025quantum,
  title={Quantum error correction below the surface code threshold},
  author={{Google Quantum AI}},
  journal={Nature},
  volume={638},
  number={8052},
  pages={920--926},
  year={2025},
  doi={10.1038/s41586-024-08449-y}
}

@article{caune2024demonstrating,
  title={Demonstrating real-time and low-latency quantum error correction with superconducting qubits},
  author={Caune, L. and Skoric, L. and Blunt, N. S. and Ruban, A. and McDaniel, J. and Valery, J. A. and Patterson, A. D. and Gramolin, A. V. and Majaniemi, J. and Barnes, K. M. and others},
  journal={arXiv:2410.05202},
  year={2024},
  url={https://doi.org/10.48550/arXiv.2410.05202}
}

@article{lacroix2025scaling,
  title={Scaling and logic in the color code on a superconducting quantum processor},
  author={Lacroix, N. and Bourassa, A. and Heras, F. J. H. and Zhang, L. M. and Bausch, J. and Senior, A. W. and Edlich, T. and Shutty, N. and Sivak, V. and Bengtsson, A. and others},
  journal={Nature},
  volume = {645},
  pages={614},
  year={2025},
  doi={10.1038/s41586-025-09061-4}
}

@article{putterman2024hardware,
  title={Hardware-efficient quantum error correction via concatenated bosonic qubits},
  author={Putterman, Harald and Noh, Kyungjoo and Hann, Connor T and MacCabe, Gregory S and Aghaeimeibodi, Shahriar and Patel, Rishi N and Lee, Menyoung and Jones, William M and Moradinejad, Hesam and Rodriguez, Roberto and others},
  journal={Nature},
  volume={638},
  number={8052},
  pages={927--934},
  year={2025},
  publisher={Nature Publishing Group UK London},
  doi={10.1038/s41586-025-08642-7}
}

@article{reichardt2024demonstration,
  title={Demonstration of quantum computation and error correction with a tesseract code},
  author={Reichardt, Ben W. and Aasen, David and Chao, Rui and Chernoguzov, Alex and van Dam, Wim and Gaebler, John P. and Gresh, Dan and Lucchetti, Dominic and Mills, Michael and Moses, Steven A. and Neyenhuis, Brian and Paetznick, Adam and Paz, Andres and Siegfried, Peter E. and da Silva, Marcus P. and Svore, Krysta M. and Wang, Zhenghan and Zanner, Matt},
  journal={arXiv:2409.04628},
  year={2024},
  url={https://doi.org/0.48550/arXiv.2409.04628}
}

@article{pogorelov2025experimental,
  title={Experimental fault-tolerant code switching},
  author={Pogorelov, I. and Butt, F. and Postler, L. and Marciniak, C. D. and Schindler, P. and M{\"u}ller, M. and Monz, T.},
  journal={Nat. Phys.},
  volume={21},
  number={2},
  pages={298--303},
  year={2025},
  doi={10.1038/s41567-024-02727-2},
  url={https://doi.org/10.1038/s41567-024-02727-2}
}

@article{bombin2006topological,
  title={Topological quantum distillation},
  author={Bombin, H. and Martin-Delgado, M. A.},
  journal={Phys. Rev. Lett.},
  volume={97},
  number={18},
  pages={180501},
  year={2006},
  doi={10.1103/PhysRevLett.97.180501},
  url={https://doi.org/10.1103/PhysRevLett.97.180501}
}

@article{daguerre2025experimental,
  title={Experimental demonstration of high-fidelity logical magic states from code switching},
  author={Daguerre, L. and Blume-Kohout, R. and Brown, N. C. and Hayes, D. and Kim, I. H.},
  journal={Phys. Rev. X},
  volume={15},
  number={4},
  pages={041008},
  year={2025},
  doi={10.1103/dck4-x9c2}
}

@article{bluvstein2024logical,
  title={Logical quantum processor based on reconfigurable atom arrays},
  author={Bluvstein, D. and Evered, S. J. and Geim, A. A. and Li, S. H. and Zhou, H. and Manovitz, T. and Ebadi, S. and Cain, M. and Kalinowski, M. and Hangleiter, D. and others},
  journal={Nature},
  volume={626},
  number={7997},
  pages={58--65},
  year={2024},
  doi={10.1038/s41586-023-06927-3},
  url={https://doi.org/10.1038/s41586-023-06927-3}
}

@article{sales2025experimental,
  title={Experimental demonstration of logical magic state distillation},
  author={Rodriguez, P. S. and Robinson, J. M. and Jepsen, P. N. and He, Z. and Duckering, C. and Zhao, C. and Wu, K.-H. and Campo, J. and Bagnall, K. and Kwon, M. and others},
  journal={Nature},
  volume={645},
  number={8081},
  pages={620--625},
  year={2025},
  doi={10.1038/s41586-025-09367-3}
}

@article{eastin2009restrictions,
  title={Restrictions on transversal encoded quantum gate sets},
  author={Eastin, B. and Knill, E.},
  journal={Phys. Rev. Lett.},
  volume={102},
  number={11},
  pages={110502},
  year={2009},
  doi={10.1103/PhysRevLett.102.110502},
  url={https://doi.org/10.1103/PhysRevLett.102.110502}
}

@article{campbell_roads_2017,
	title = {Roads towards fault-tolerant universal quantum computation},
	volume = {549},
	copyright = {2017 Macmillan Publishers Limited, part of Springer Nature. All rights reserved.},
	issn = {1476-4687},
	url = {https://www.nature.com/articles/nature23460},
	doi = {10.1038/nature23460},
	number = {7671},
	urldate = {2026-05-22},
	journal = {Nature},
	publisher = {Nature Publishing Group},
	author = {Campbell, Earl T. and Terhal, Barbara M. and Vuillot, Christophe},
	month = sep,
	year = {2017},
	pages = {172--179},
}

@article{chou2018deterministic,
	title = {Deterministic teleportation of a quantum gate between two logical qubits},
	volume = {561},
	copyright = {2018 Springer Nature Limited},
	issn = {1476-4687},
	url = {https://www.nature.com/articles/s41586-018-0470-y},
	doi = {10.1038/s41586-018-0470-y},
	number = {7723},
	urldate = {2026-05-22},
	journal = {Nature},
	publisher = {Nature Publishing Group},
	author = {Chou, Kevin S. and Blumoff, Jacob Z. and Wang, Christopher S. and Reinhold, Philip C. and Axline, Christopher J. and Gao, Yvonne Y. and Frunzio, L. and Devoret, M. H. and Jiang, Liang and Schoelkopf, R. J.},
	month = sep,
	year = {2018},
	pages = {368--373},
}

@article{gottesman1999demonstrating,
	title = {Demonstrating the viability of universal quantum computation using teleportation and single-qubit operations},
	volume = {402},
	copyright = {1999 Macmillan Magazines Ltd.},
	issn = {1476-4687},
	url = {https://www.nature.com/articles/46503},
	doi = {10.1038/46503},
	number = {6760},
	urldate = {2026-05-22},
	journal = {Nature},
	publisher = {Nature Publishing Group},
	author = {Gottesman, Daniel and Chuang, Isaac L.},
	month = nov,
	year = {1999},
	pages = {390--393},
}

@article{bravyi2005universal,
  title={Universal quantum computation with ideal {C}lifford gates and noisy ancillas},
  author={Bravyi, S. and Kitaev, A.},
  journal={Phys. Rev. A},
  volume={71},
  number={2},
  pages={022316},
  year={2005},
  doi={10.1103/PhysRevA.71.022316},
  url={https://doi.org/10.1103/PhysRevA.71.022316}
}

@article{gidney2024magic,
  title={Magic state cultivation: Growing {T} states as cheap as {CNOT} gates},
  author={Gidney, C. and Shutty, N. and Jones, C.},
  journal={arXiv:2409.17595},
  year={2024},
  url={https://doi.org/10.48550/arXiv.2409.17595}
}

@article{rosenfeld2025magic,
  title={Magic state cultivation on a superconducting quantum processor},
  author={Rosenfeld, E. and Gidney, C. and Roberts, G. and Morvan, A. and Lacroix, N. and Kafri, D. and Marshall, J. and Li, M. and Sivak, V. and Abanin, D. and others},
  journal={arXiv:2512.13908},
  year={2025},
  url={https://doi.org/10.48550/arXiv.2512.13908}
}

@article{bombin2015gauge,
  title={Gauge color codes: optimal transversal gates and gauge fixing in topological stabilizer codes},
  author={Bombin, H.},
  journal={New J. Phys.},
  volume={17},
  number={8},
  pages={083002},
  year={2015},
  doi={10.1088/1367-2630/17/8/083002},
  url={https://doi.org/10.1088/1367-2630/17/8/083002}
}

@article{anderson2014fault,
  title={Fault-tolerant conversion between the {S}teane and {R}eed-{M}uller quantum codes},
  author={Anderson, J. T. and Duclos-Cianci, G. and Poulin, D.},
  journal={Phys. Rev. Lett.},
  volume={113},
  number={8},
  pages={080501},
  year={2014},
  doi={10.1103/PhysRevLett.113.080501},
  url={https://doi.org/10.1103/PhysRevLett.113.080501}
}

@article{kubica2015universal,
  title={Universal transversal gates with color codes: A simplified approach},
  author={Kubica, A. and Beverland, M. E.},
  journal={Phys. Rev. A},
  volume={91},
  number={3},
  pages={032330},
  year={2015},
  doi={10.1103/PhysRevA.91.032330},
  url={https://doi.org/10.1103/PhysRevA.91.032330}
}

@article{yamamoto2026quantum,
	title = {Quantum {Error}-{Corrected} {Computation} of {Molecular} {Energies}},
	volume = {7},
	url = {https://link.aps.org/doi/10.1103/m7j3-5sk6},
	doi = {10.1103/m7j3-5sk6},
	number = {2},
	urldate = {2026-05-22},
	journal = {PRX Quantum},
	publisher = {American Physical Society},
	author = {Yamamoto, Kentaro and Kikuchi, Yuta and Amaro, David and Criger, Ben and Dilkes, Silas and Ryan-Anderson, Ciarán and Tranter, Andrew and Dreiling, Joan M. and Gresh, Dan and Foltz, Cameron and Mills, Michael and Moses, Steven A. and Siegfried, Peter E. and Urmey, Maxwell D. and Burau, Justin J. and Hankin, Aaron and Lucchetti, Dominic and Gaebler, John P. and Brown, Natalie C. and Neyenhuis, Brian and Ramo, David Muñoz},
	month = apr,
	year = {2026},
	pages = {020319},
}

@article{butt2024fault,
  title={Fault-tolerant code-switching protocols for near-term quantum processors},
  author={Butt, F. and Heu{\ss}en, S. and Rispler, M. and M{\"u}ller, M.},
  journal={PRX Quantum},
  volume={5},
  number={2},
  pages={020345},
  year={2024},
  doi={10.1103/PRXQuantum.5.020345},
  url={https://doi.org/10.1103/PRXQuantum.5.020345}
}

@article{bombin2007topological,
  title={Topological computation without braiding},
  author={Bombin, Hector and Martin-Delgado, Miguel-Angel},
  journal={Phys. Rev. Lett.},
  volume={98},
  number={16},
  pages={160502},
  year={2007},
  publisher={APS},
  doi={10.1103/PhysRevLett.98.160502}
}

@article{bombin2013self,
  title={Self-correcting quantum computers},
  author={Bombin, Hector and Chhajlany, Ravindra W and Horodecki, Micha{\l} and Martin-Delgado, Miguel-Angel},
  journal={New Journ. Phys.},
  volume={15},
  number={5},
  pages={055023},
  year={2013},
  publisher={IOP Publishing},
doi={10.1088/1367-2630/15/5/055023}
}

@article{huang2024comparing,
  title={Comparing {S}hor and {S}teane error correction using the {B}acon-{S}hor code},
  author={Huang, S. and Brown, K. R. and Cetina, M.},
  journal={Sci. Adv.},
  volume={10},
  number={45},
  pages={eadp2008},
  year={2024},
  doi={10.1126/sciadv.adp2008}
}

@article{yoder2016universal,
  title={Universal fault-tolerant gates on concatenated stabilizer codes},
  author={Yoder, Theodore J and Takagi, Ryuji and Chuang, Isaac L},
  journal={Phys. Rev. X},
  volume={6},
  number={3},
  pages={031039},
  year={2016},
  publisher={APS},
  doi={10.1103/PhysRevX.6.031039}
}

@article{ryan2022implementing,
  title={Implementing fault-tolerant entangling gates on the five-qubit code and the color code},
  author={Ryan-Anderson, C and Brown, NC and Allman, MS and Arkin, B and Asa-Attuah, G and Baldwin, C and Berg, J and Bohnet, JG and Braxton, S and Burdick, N and others},
  journal={arXiv:2208.01863},
  year={2022},
  url={https://doi.org/10.48550/arXiv.2208.01863}
}

@article{heussen2025efficient,
  title={Efficient fault-tolerant code switching via one-way transversal CNOT gates},
  author={Heu{\ss}en, Sascha and Hilder, Janine},
  journal={Quantum},
  volume={9},
  pages={1846},
  year={2025},
  publisher={Verein zur F{\"o}rderung des Open Access Publizierens in den Quantenwissenschaften},
  doi={10.22331/q-2025-09-03-1846}
}

@article{tan2025single,
  title={Single-shot universality in quantum {LDPC} codes via code-switching},
  author={Tan, Shi Jie Samuel and Hong, Yifan and Lin, Ting-Chun and Gullans, Michael J and Hsieh, Min-Hsiu},
  journal={arXiv:2510.08552},
  year={2025},
  url={https://doi.org/10.48550/arXiv.2510.08552}
}

@article{hangleiter2025fault,
  title={Fault-tolerant compiling of classically hard instantaneous quantum polynomial circuits on hypercubes},
  author={Hangleiter, Dominik and Kalinowski, Marcin and Bluvstein, Dolev and Cain, Madelyn and Maskara, Nishad and Gao, Xun and Kubica, Aleksander and Lukin, Mikhail D and Gullans, Michael J},
  journal={PRX Quantum},
  volume={6},
  number={2},
  pages={020338},
  year={2025},
  publisher={APS},
  doi={10.1103/PRXQuantum.6.02033}
}

@article{yoder2017universal,
  title={Universal fault-tolerant quantum computation with {B}acon-{S}hor codes},
  author={Yoder, Theodore J},
  journal={arXiv:1705.01686},
  year={2017},
url={https://doi.org/10.48550/arXiv.1705.01686}
}

@article{lin2020concatenated,
  title={Concatenated pieceable fault-tolerant scheme for universal quantum computation},
  author={Lin, Chen and Yang, GuoWu},
  journal={Phys. Rev. A},
  volume={102},
  number={5},
  pages={052415},
  year={2020},
  publisher={APS},
  doi={10.1103/PhysRevA.102.052415}
}

@article{takagi2017error,
  title={Error rates and resource overheads of encoded three-qubit gates},
  author={Takagi, Ryuji and Yoder, Theodore J and Chuang, Isaac L},
  journal={Phys. Rev. A},
  volume={96},
  number={4},
  pages={042302},
  year={2017},
  publisher={APS},
doi={10.1103/PhysRevA.96.042302}
}

@article{chao2018fault,
  title={Fault-tolerant quantum computation with few qubits},
  author={Chao, Rui and Reichardt, Ben W},
  journal={npj Quant. Inf.},
  volume={4},
  number={1},
  pages={42},
  year={2018},
  publisher={Nature Publishing Group UK London},
doi={10.1038/s41534-018-0085-z}
}

@article{jochym2016stacked,
  title={Stacked codes: Universal fault-tolerant quantum computation in a two-dimensional layout},
  author={Jochym-O'Connor, Tomas and Bartlett, Stephen D},
  journal={Phys. Rev. A},
  volume={93},
  number={2},
  pages={022323},
  year={2016},
  publisher={APS},
doi={10.1103/PhysRevA.93.022323}
}

@article{steane1997active,
  title={Active stabilization, quantum computation, and quantum state synthesis},
  author={Steane, Andrew M},
  journal={Phys. Rev. Lett.},
  volume={78},
  number={11},
  pages={2252},
  year={1997},
  publisher={APS},
  doi={10.1103/PhysRevLett.78.2252}
}

@article{reichardt2024fault,
  title={Fault-tolerant quantum computation with a neutral atom processor},
  author={Reichardt, Ben W and Paetznick, Adam and Aasen, David and Basov, Ivan and Bello-Rivas, Juan M and Bonderson, Parsa and Chao, Rui and van Dam, Wim and Hastings, Matthew B and Mishmash, Ryan V and others},
  journal={arXiv:2411.11822},
  year={2024},
url={https://doi.org/10.48550/arXiv.2411.11822}
}

@article{butt2026decoding,
  title={Decoding three-dimensional color codes with boundaries},
  author={Butt, Friederike and Esser, Lars and M{\"u}ller, Markus},
  journal={Phys. Rev. A},
  volume={113},
  number={4},
  pages={042416},
  year={2026},
  publisher={APS},
doi={10.1103/7cdb-px1d}
}

@article{butt2026demonstration,
  title        = {Demonstration of measurement-free universal logical quantum computation},
  author       = {Butt, F. and Pogorelov, I. and Freund, R. and Steiner, A. and Meyer, M. and Monz, T. and M{\"u}ller, M.},
  journal 	   = {Nat. Commun.},
  volume       = {17},
  pages        = {995},
  year         = {2026},
  doi          = {10.1038/s41467-026-68533-x}
}

@article{mayer2024benchmarking,
  title={Benchmarking logical three-qubit quantum {F}ourier transform encoded in the {S}teane code on a trapped-ion quantum computer},
  author={Mayer, K. and Ryan-Anderson, C. and Brown, N. and Durso-Sabina, E. and Baldwin, C. H. and Hayes, D. and Dreiling, J. M. and Foltz, C. and Gaebler, J. P. and Gatterman, T. M. and others},
  journal={arXiv:2404.08616},
  year={2024},
  url={https://doi.org/10.48550/arXiv.2404.08616}
}

@article{postler2022demonstration,
  title={Demonstration of fault-tolerant universal quantum gate operations},
  author={Postler, L. and Heu{\ss}en, S. and Pogorelov, I. and Rispler, M. and Feldker, T. and Meth, M. and Marciniak, C. D. and Stricker, R. and Ringbauer, M. and Blatt, R. and others},
  journal={Nature},
  volume={605},
  number={7911},
  pages={675--680},
  year={2022},
  doi={10.1038/s41586-022-04721-1},
  url={https://doi.org/10.1038/s41586-022-04721-1}
}

@article{postler2024demonstration,
  title={Demonstration of fault-tolerant {S}teane quantum error correction},
  author={Postler, L. and Butt, F. and Pogorelov, I. and Marciniak, C. D. and Heu{\ss}en, S. and Blatt, R. and Schindler, P. and Rispler, M. and M{\"u}ller, M. and Monz, T.},
  journal={PRX Quantum},
  volume={5},
  number={3},
  pages={030326},
  year={2024},
  doi={10.1103/PRXQuantum.5.030326}
}

@article{sahay2025fold,
  title={Fold-transversal surface code cultivation},
  author={Sahay, K. and Tsai, P.-K. and Chang, K. and Su, Q. and Smith, T. B. and Singh, S. and Puri, S.},
  journal={arXiv:2509.05212},
  year={2025},
  url={https://doi.org/10.48550/arXiv.2509.05212}
}

@article{evered2023high,
  title={High-fidelity parallel entangling gates on a neutral-atom quantum computer},
  author={Evered, Simon J and Bluvstein, Dolev and Kalinowski, Marcin and Ebadi, Sepehr and Manovitz, Tom and Zhou, Hengyun and Li, Sophie H and Geim, Alexandra A and Wang, Tout T and Maskara, Nishad and others},
  journal={Nature},
  volume={622},
  number={7982},
  pages={268--272},
  year={2023},
  publisher={Nature Publishing Group UK London},
doi={10.1038/s41586-023-06481-y}
}

@article{haffner2008quantum,
  title={Quantum computing with trapped ions},
  author={H{\"a}ffner, Hartmut and Roos, Christian F and Blatt, Rainer},
  journal={Phys. Rep.},
  volume={469},
  number={4},
  pages={155--203},
  year={2008},
  publisher={Elsevier},
doi={10.1016/j.physrep.2008.09.003}
}

@article{preskill1998reliable,
  title={Reliable quantum computers},
  author={Preskill, John},
  journal={Proceedings of the Royal Society of London. Series A: Mathematical, Physical and Engineering Sciences},
  volume={454},
  number={1969},
  pages={385--410},
  year={1998},
  publisher={The Royal Society},
doi={10.1098/rspa.1998.0167}
}

@article{bluvstein2025fault,
  title={A fault-tolerant neutral-atom architecture for universal quantum computation},
  author={Bluvstein, D. and Geim, A. A. and Li, S. H. and Evered, S. J. and Bonilla Ataides, J. P. and Baranes, G. and Gu, A. and Manovitz, T. and Xu, M. and Kalinowski, M. and others},
  journal={Nature},
  volume={649},
  pages={39--46},
  year={2026},
  doi={10.1038/s41586-025-09848-5}
}

@article{kitaev1997quantum,
  title={Quantum computations: algorithms and error correction},
  author={Kitaev, A. Yu.},
  journal={Russ. Math. Surv.},
  volume={52},
  number={6},
  pages={1191},
  year={1997},
  doi={10.1070/RM1997v052n06ABEH002155},
  url={https://doi.org/10.1070/RM1997v052n06ABEH002155}
}

@article{dawson2005solovay,
  title={The {S}olovay-{K}itaev algorithm},
  author={Dawson, C. M. and Nielsen, M. A.},
  journal={arXiv:quant-ph/0505030},
  year={2005},
  url={https://doi.org/10.48550/arXiv.quant-ph/0505030}
}

@article{mathiot2026benchmarking,
  title={Benchmarking a machine-learning differential equations solver on a neutral-atom logical processor},
  author={Mathiot, Pauline and Garnaoui, Elio and Leriche, Axel-Ugo and Philip, Evan and Albrecht, Boris and Briosne-Fr{\'e}javille, Cl{\'e}mence and Cardarelli, Lorenzo and Cornillot, Antoine and Cournez, Gwennol{\'e} and Couturier, Luc and others},
  journal={arXiv:2605.21276},
  year={2026},
url={https://doi.org/10.48550/arXiv.2605.21276}
}

@article{campbell2017roads,
  title={Roads towards fault-tolerant universal quantum computation},
  author={Campbell, Earl T and Terhal, Barbara M and Vuillot, Christophe},
  journal={Nature},
  volume={549},
  number={7671},
  pages={172--179},
  year={2017},
  publisher={Nature Publishing Group UK London},
  doi={10.1038/nature23460}
}

@article{aliferis2005quantum,
  title={Quantum accuracy threshold for concatenated distance-3 codes},
  author={Aliferis, Panos and Gottesman, Daniel and Preskill, John},
  journal={arXiv quant-ph/0504218},
  year={2005},
  url = {https://doi.org/10.48550/arXiv.quant-ph/0504218}
}

@article{gutierrez2015comparison,
  title={Comparison of a quantum error-correction threshold for exact and approximate errors},
  author={Guti{\'e}rrez, Mauricio and Brown, Kenneth R},
  journal={Phys. Rev. A},
  volume={91},
  pages={022335},
  year={2015},
  publisher={APS},
  doi={https://doi.org/10.1103/PhysRevA.91.022335}
}

@article{cross2007comparative,
author = {Cross, Andrew W. and Divincenzo, David P. and Terhal, Barbara M.},
title = {A comparative code study for quantum fault tolerance},
year = {2009},
issue_date = {July 2009},
publisher = {Rinton Press, Incorporated},
address = {Paramus, NJ},
volume = {9},
issn = {1533-7146},
journal = {Quantum Info. Comput.},
month = jul,
pages = {541–572},
numpages = {32},
doi={10.48550/arXiv.0711.1556}
}

@article{landahl2011fault,
  title={Fault-tolerant quantum computing with color codes},
  author={Landahl, Andrew J and Anderson, Jonas T and Rice, Patrick R},
  journal={arXiv:1108.5738},
  year={2011},
  url={https://doi.org/10.48550/arXiv.1108.5738}
}

@article{goto2016minimizing,
  title={Minimizing resource overheads for fault-tolerant preparation of encoded states of the {S}teane code},
  author={Goto, H.},
  journal={Sci. Rep.},
  volume={6},
  number={1},
  pages={19578},
  year={2016},
  doi={10.1038/srep19578},
  url={https://doi.org/10.1038/srep19578}
}

@article{howard2017application,
  title={Application of a resource theory for magic states to fault-tolerant quantum computing},
  author={Howard, Mark and Campbell, Earl},
  journal={Phys. Rev. Lett.},
  volume={118},
  number={9},
  pages={090501},
  year={2017},
  publisher={APS},
  doi={10.1103/PhysRevLett.118.090501}
}

@article{kubica2015unfolding,
  title={Unfolding the color code},
  author={Kubica, A. and Yoshida, B. and Pastawski, F.},
  journal={New J. Phys.},
  volume={17},
  number={8},
  pages={083026},
  year={2015},
  doi={10.1088/1367-2630/17/8/083026}
}

@article{zhou2025low,
  title={Low-overhead transversal fault tolerance for universal quantum computation},
  author={Zhou, Hengyun and Zhao, Chen and Cain, Madelyn and Bluvstein, Dolev and Maskara, Nishad and Duckering, Casey and Hu, Hong-Ye and Wang, Sheng-Tao and Kubica, Aleksander and Lukin, Mikhail D},
  journal={Nature},
  volume={646},
  number={8084},
  pages={303--308},
  year={2025},
  publisher={Nature Publishing Group UK London},
  doi={10.1038/s41586-025-09543-5}
}

@article{serra2026decoding,
  title={Decoding across transversal {C}lifford gates in the surface code},
  author={Serra-Peralta, Marc and Shaw, Mackenzie H and Terhal, Barbara M},
  journal={PRX Quantum},
  volume={7},
  number={1},
  pages={010335},
  year={2026},
  publisher={APS},
  doi={10.1103/sk5y-25b1}
}

@article{shepherd2009temporally,
  title={Temporally unstructured quantum computation},
  author={Shepherd, Dan and Bremner, Michael J},
  journal={Proceedings of the Royal Society A: Mathematical, Physical and Engineering Sciences},
  volume={465},
  number={2105},
  pages={1413--1439},
  year={2009},
  publisher={The Royal Society London},
  doi={10.1098/rspa.2008.0443}
}

@article{paletta2024robust,
  title={Robust sparse IQP sampling in constant depth},
  author={Paletta, Louis and Leverrier, Anthony and Sarlette, Alain and Mirrahimi, Mazyar and Vuillot, Christophe},
  journal={Quantum},
  volume={8},
  pages={1337},
  year={2024},
  publisher={Verein zur F{\"o}rderung des Open Access Publizierens in den Quantenwissenschaften},
  doi={10.22331/q-2024-05-06-1337}
}

@article{bremner2017achieving,
  title={Achieving quantum supremacy with sparse and noisy commuting quantum computations},
  author={Bremner, Michael J and Montanaro, Ashley and Shepherd, Dan J},
  journal={Quantum},
  volume={1},
  pages={8},
  year={2017},
  publisher={Verein zur F{\"o}rderung des Open Access Publizierens in den Quantenwissenschaften},
  doi={10.22331/q-2017-04-25-8}
}

\end{document}